\documentclass[epjCONF]{svjour}
\usepackage{graphicx}
\usepackage{epsfig}
\usepackage[varg]{txfonts} 
\usepackage[latin1]{inputenc}
\session-title{Search for the He-$\eta$ bound states with the WASA-at-COSY facility}
\begin{document}
\title{Search for the He-$\eta$ bound states with the WASA-at-COSY facility}
\author{M. Skurzok\inst{1}\fnmsep\thanks{\email{mskurzok@gmail.com}} \and W. Krzemien \inst{1} \and P. Moskal \inst{1,2}}
\institute{M. Smoluchowski Institute of Physics, Jagiellonian University, Cracow, Poland \and Institut fur Kernphysik, Forschungszentrum J\"ulich, Germany}

\abstract{
The $\eta$-mesic nuclei in which the $\eta$ meson is bound with nucleus via strong interaction was postulated already in 1986, however till now no experiment confirmed empirically its existence. The discovery of this new kind of an exotic nuclear matter would be very important for better understanding of the $\eta$ meson structure and its interaction with nucleons. 
The search for $\eta$-mesic helium is carried out with high statistic and high acceptance with the WASA-at-COSY detection setup in the Research Center J\"ulich.
The search is conducted via the measurement of the excitation function for the chosen decay channels of the $^{4}\hspace{-0.03cm}\mbox{He}$-$\eta$ system. 
Till now two reactions $dd\rightarrow(^{4}\hspace{-0.03cm}\mbox{He}$-$\eta)_{bs}\rightarrow$ $^{3}\hspace{-0.03cm}\mbox{He} p \pi{}^{-}$ and \mbox{$dd\rightarrow(^{4}\hspace{-0.03cm}\mbox{He}$-$\eta)_{bs}\rightarrow$ $^{3}\hspace{-0.03cm}\mbox{He} n \pi{}^{0}$} were measured with the beam momentum ramped around the $\eta$ production threshold. This report includes the description of experimental method and status of the analysis.
} 

\maketitle
\section{Introduction}
\label{intro}

Based on the fact that the interaction between the $\eta$ meson and nucleon
is attractive Haider and Liu to postulate the existence of the $\eta$-mesic nuclei~\cite{HaiderLiu1},
in which the neutral $\eta$ meson might be bound with nucleons via the strong
interaction. The existence of $\eta$-mesic bound states could allow to investigate the $\eta$ meson properties~\cite{InoueOset} and interaction between the $\eta$ meson and nucleons inside a nuclear matter.~Furthermore it would provide information about the $\mbox{N}^{*}(1535)$ resonance~\cite{Jido}, as well as about contribution of the flavour singlet component of the quark-gluon wave function of the $\eta$ meson~\cite{BassTom}. 
According to the theoretical considerations, the formation of the $\eta$-mesic nucleus can only take place if the real part of the $\eta$-nucleus scattering length is negative (attractive interaction), and the magnitude of the real part is greater than the magnitude of the imaginary part~\cite{HaiderLiu2}:

\begin{equation}
|Re(a_{\eta-nucleus})|>|Im(a_{\eta-nucleus})|.\label{eq:eq1}
\end{equation}

\noindent Calculations for hadronic- and photoproduction of the $\eta$ meson gave a wide range of possible values of the s-wave $\eta$N scattering lenght, from $a_{\eta N}$=(0.27 + 0.22i) fm up to $a_{\eta N}$=(1.05 + 0.27i) fm. Such a high values has not exluded the formation of $\eta$-nucleus bound states for a light nuclei as $^{3,4}\hspace{-0.03cm}\mbox{He}$, T~\cite{Wilkin1,WycechGreen} and even for deuteron~\cite{Green}. 
Those bound states have been searched in many experiments~\cite{Machner,Sokol1,Gillitzer,Berger,Mayer,Mersmann,Smyrski1}, however none of them gave empirical confirmation of their existence. There are only a promissing experimental signals which might be interpreted as~\mbox{indications} of the $\eta$-mesic nuclei. For example, experimental observations which might suggest the possibility of~the existence of the $^{3}\hspace{-0.03cm}\mbox{He}$-$\eta$ bound system were found by \mbox{SPES-4}~\cite{Berger}, \mbox{SPES-2}~\cite{Mayer}, ANKE~\cite{Mersmann}, \mbox{COSY-11}~\cite{Smyrski1} and TAPS~\cite{Pfeiffer} collaborations. 

\section{Experiment}

\begin{figure}[h]
\centering
\includegraphics[width=9.5cm,height=5.0cm]{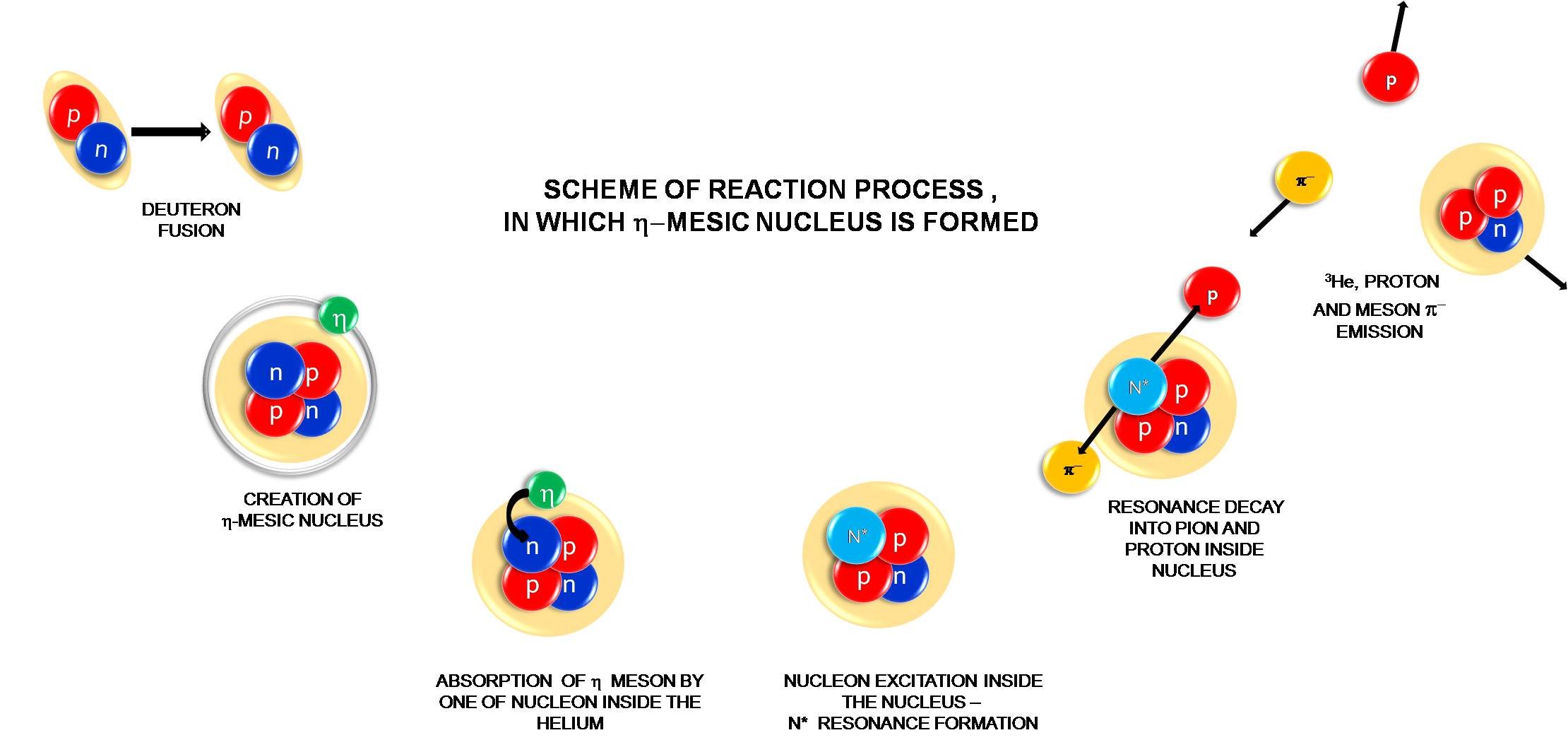}
\caption{Reaction process of the ($^{4}\hspace{-0.03cm}\mbox{He}$-$\eta)_{bs}$ production and decay.}
\label{fig1}
\end{figure}

The measurement of the $^{4}\hspace{-0.03cm}\mbox{He}$-$\eta$ bound states is carried out with unique precision  by means of the WASA detection system, installed at the COSY synchrotron.~Signals of the \mbox{$\eta$-mesic} nuclei are searched for via studying the excitation function of specific decay channels of the \mbox{$^{4}\hspace{-0.03cm}\mbox{He}$-$\eta$} system, formed in deuteron-deuteron collision~\cite{Moskal1,Krzemien}.~The measurement is performed for the beam momenta varying continously around the $\eta$ production threshold.~The~beam ramping technique allows to reduce the systematic uncertainities.~The \mbox{existence} of the bound system should be manifested as a resonance-like structure in the excitation curve of eg. $dd\rightarrow(^{4}\hspace{-0.03cm}\mbox{He}$-$\eta)_{bs}\rightarrow$ $^{3}\hspace{-0.03cm}\mbox{He} p \pi{}^{-}$ reaction
below the $dd\rightarrow$ $^{4}\hspace{-0.03cm}\mbox{He}$-$\eta$ reaction threshold. 
The reaction kinematics is schematically presented in Fig.~\ref{fig1}.


The deuteron - deuteron collision leads to the creation of the $^{4}\hspace{-0.03cm}\mbox{He}$-$\eta$ bound system. The $\eta$ meson can be absorbed by one of the nucleons inside helium and may propagate in the nucleus via consecutive excitation of nucleons to the $\mbox{N}^{*}(1525)$ state~\cite{Sokol} until the resonance decays into the pion-proton pair outgoing from the nucleus~\cite{KrzeMosSmy}. The relative angle between p and $\pi^{-}$ is equal to
$180^\circ$ in the  $\mbox{N}^{*}$ reference frame and it is smeared by about $30^\circ$ in the center-of-mass frame due to the Fermi motion of the nucleons inside the helium nucleus.

In June 2008 a search for the $^{4}\hspace{-0.03cm}\mbox{He}$-$\eta$ bound state was performed by measuring the excitation function of the $dd\rightarrow$ $^{3}\hspace{-0.03cm}\mbox{He} p \pi{}^{-}$ reaction near the $\eta$ production threshold.~During the experiment the deuteron beam momentum was varied continuously from 2.185~GeV/c to 2.400~GeV/c what corresponds the excess energy variation from -51.4 MeV to 22 MeV. 
Excitation function was determined after applying cuts on the p and $\pi^{-}$ kinetic energy distribution and the $p - \pi^{-}$ opening angle in the CM system~\cite{Krzemien_PhD}. The relative normalization of points of the $dd\rightarrow$ $^{3}\hspace{-0.03cm}\mbox{He} p \pi{}^{-}$ excitation function was based on the quasi-elastic proton-proton scattering.
In the excitation function there is no structure which could be interpreted as a resonance originating from decay of the $\eta$-mesic $^{4}\hspace{-0.03cm}\mbox{He}$~\cite{Krzemien_MESON2012}.

During the experiment, in November 2010, two channels of the $\eta$-mesic helium decay were measured:  $dd\rightarrow(^{4}\hspace{-0.03cm}\mbox{He}$-$\eta)_{bs}\rightarrow$ $^{3}\hspace{-0.03cm}\mbox{He} p \pi{}^{-}$ and  $dd\rightarrow(^{4}\hspace{-0.03cm}\mbox{He}$-$\eta)_{bs}\rightarrow$ $^{3}\hspace{-0.03cm}\mbox{He} n \pi{}^{0} \rightarrow$ $^{3}\hspace{-0.03cm}\mbox{He} n \gamma \gamma$~\cite{MSkurzok}. The measurement was performed with the beam momentum ramping from 2.127GeV/c to 2.422GeV/c, corresponding to the range of the excess energy \mbox{Q$\in$(-70,30)~MeV}. Data were effectively taken for about~155 hours. The average luminosity was estimated based on the binary reaction \mbox{$dd\rightarrow$ $^{3}\hspace{-0.03cm}\mbox{He} n$} and is equal to about L=1.5$\cdot10^{31} cm^{-2} s^{-1}$. Taking into account the fact that two reactions were measured, in total more than 40 times higher statistics were collected than in the experiment carried out in 2008. At present the data analysis is in progress. In the optimistic case, the statistics could be sufficient to observe a signal from the $\eta$-mesic helium and in the pesimistic scenario the upper limit of the cross section for the $^{4}\hspace{-0.03cm}\mbox{He}$-$\eta$ bound state production will be decreased by a factor of about six. 


\section{Acknowledgements}
\noindent We acknowledge support by the Foundation for Polish Science - MPD program, co-financed by the European
Union within the European Regional Development Fund, by the Polish National Science Center through grant No. 2011/01/B/ST2/00431 and by the FFE grants of the Research Center Juelich.

\label{sec:1}
\label{sec:2}
\label{sec:3}

%

%

\end{document}